\documentclass[sigconf,screen]{acmart}

\usepackage{multirow}
\usepackage{makecell}
\usepackage{pifont}

\usepackage{algorithm}
\usepackage{algpseudocode}

\AtBeginDocument{}

\copyrightyear{2026}
\acmYear{2026}
\setcopyright{cc}
\setcctype{by}
\acmConference[ICSE '26]{2026 IEEE/ACM 48th International Conference on Software Engineering}{April 12--18, 2026}{Rio de Janeiro, Brazil}
\acmBooktitle{2026 IEEE/ACM 48th International Conference on Software Engineering (ICSE '26), April 12--18, 2026, Rio de Janeiro, Brazil}
\acmPrice{}
\acmDOI{10.1145/3744916.3773242}
\acmISBN{979-8-4007-2025-3/26/04}

\RequirePackage[normalem]{ulem} 
\RequirePackage{color}\definecolor{myadd}{RGB}{0,0,0}\definecolor{mydel}{RGB}{0,0,0} 
\providecommand{\DIFadd}[1]{\protect{\color{myadd}#1}} 
\providecommand{\DIFaddbegin}{} 
\providecommand{\DIFaddend}{} 
\providecommand{\DIFdelbegin}{} 
\providecommand{\DIFdelend}{} 
\providecommand{\DIFaddFL}[1]{\DIFadd{#1}} 
\providecommand{\DIFaddbeginFL}{} 
\providecommand{\DIFaddendFL}{} 
\providecommand{\DIFdelbeginFL}{} 
\providecommand{\DIFdelendFL}{} 
\RequirePackage{listings} 
\RequirePackage{color} 
\lstdefinelanguage{DIFcode}{ 
  moredelim=[il][\color{red}\sout]{\%DIF\ <\ }, 
  moredelim=[il][\color{blue}\uwave]{\%DIF\ >\ } 
} 
\lstdefinestyle{DIFverbatimstyle}{ 
	language=DIFcode, 
	basicstyle=\ttfamily, 
	columns=fullflexible, 
	keepspaces=true 
} 
\lstnewenvironment{DIFverbatim}{\lstset{style=DIFverbatimstyle}}{} 
\lstnewenvironment{DIFverbatim*}{\lstset{style=DIFverbatimstyle,showspaces=true}}{} 

\begin{document}

\title{\DIFaddbegin \DIFadd{Scalpel: Automotive Deep Learning Framework Testing via Assembling Model Components}\DIFaddend }

\author{Yinglong Zou}
\affiliation{%
  \institution{State Key Laboratory for Novel Software Technology\\ Nanjing University}
  \country{China}
}
\email{652023320004@smail.nju.edu.cn}
\orcid{0009-0006-9375-7417}

\author{Juan Zhai}
\affiliation{%
  \institution{University of Massachusetts Amherst 
  \country{United States}}
}
\email{juanzhai@umass.edu}
\orcid{0000-0001-5017-8016}

\author{Chunrong Fang}
\authornote{Chunrong Fang and Zhenyu Chen are the corresponding authors.}
\affiliation{%
  \institution{State Key Laboratory for Novel Software Technology\\ Nanjing University}
  \country{China}
}
\email{fangchunrong@nju.edu.cn}
\orcid{0000-0002-9930-7111}

\author{An Guo}
\affiliation{%
  \institution{Hong Kong Polytechnic University}
  \country{China}
}
\email{an.guo@polyu.edu.hk}
\orcid{0009-0005-8661-6133}

\author{Jiawei Liu}
\affiliation{%
  \institution{State Key Laboratory for Novel Software Technology\\ Nanjing University}
  \country{China}
}
\email{jw.liu@nju.edu.cn}
\orcid{0000-0002-4930-9637}

\author{Zhenyu Chen}
\affiliation{%
  \institution{State Key Laboratory for Novel Software Technology\\ Nanjing University}
  \country{China}
}
\email{zychen@nju.edu.cn}
\orcid{0000-0002-9592-7022}

\renewcommand{\shortauthors}{Zou et al.}

\begin{abstract}

\DIFaddend Deep learning (DL) plays a key role in autonomous driving systems. DL models support perception modules, equipped with tasks such as object detection and sensor fusion. These DL models enable vehicles to process multi-sensor inputs to understand complex surroundings. Deploying DL models in autonomous driving systems faces stringent challenges, including real-time processing, limited computational resources, and strict power constraints. To address these challenges, automotive DL frameworks (e.g., PaddleInference) have emerged \DIFadd{to optimize inference efficiency. However,} these frameworks encounter unique quality issues due to their more complex deployment environments, such as crashes stemming from limited scheduled memory and incorrect memory allocation. Unfortunately, existing DL framework testing methods fail to detect these quality issues due to the failure in deploying generated test input models, as these models lack three essential capabilities: (1) multi-input/output tensor processing, (2) multi-modal data processing, and (3) multi-level data feature extraction. These capabilities necessitate specialized model components, which existing testing methods neglect during model generation. \DIFaddbegin \DIFadd{To bridge this gap, we propose Scalpel, an automotive DL frameworks testing method that generates test input models at the model component level. Scalpel generates models }\DIFaddend by assembling model components (heads, necks, backbones) to support capabilities required by autonomous driving systems.  
{\DIFadd{Specifically, Scalpel maintains and updates a repository of model components, generating test inputs by selecting, mutating, and assembling them. Successfully generated models are added back to enrich the repository.}
Newly generated models are then deployed within the autonomous driving system to test automotive DL frameworks via differential testing. }\DIFaddend The experimental results demonstrate that Scalpel outperforms existing methods in both effectiveness and efficiency.  \DIFaddbegin \DIFadd{In Apollo, Scalpel detects 16 crashes and 21 NaN \& inconsistency bugs. All detected bugs have been reported to open-source communities, with 10 crashes confirmed. Scalpel achieves 27.44$\times$ and 8.5$\times$ improvements in model generation efficiency and bug detection efficiency. Additionally, Scalpel detects nine crashes and 16 NaN \& inconsistency bugs in Autoware, which shows its excellent generalization. 
}\DIFaddend \end{abstract}

\maketitle
\section{Introduction}
\label{introduction}

Autonomous driving systems have rapidly evolved as a cornerstone of modern transportation innovation \cite{ADS1,ADS2,DL4ADS1}. In autonomous driving systems, deep learning (DL) has become a significant technology, particularly in perception modules where it enables critical functions such as object detection \cite{ADSPerception1,ADSPerception2} and sensor fusion \cite{ADSSensorFusion1,ADSSensorFusion2}. These DL-driven modules allow vehicles to interpret complex environments through camera and LiDAR inputs \cite{Apollo}.

The deployment of automotive DL models faces unique constraints. Autonomous driving systems must balance stringent requirements for real-time processing with limited computational resources and strict power budgets. These constraints have driven the emergence of automotive DL frameworks (e.g., PaddleInference) \DIFadd{to optimize inference efficiency}. Unlike general DL frameworks such as TensorFlow \cite{tensorflow} and PyTorch \cite{pytorch}, automotive DL frameworks face more complex deployment environments, leading to unique quality issues. For example, resource constraints in automotive DL frameworks often trigger critical bugs that rarely occur in general DL frameworks, such as crashes caused by constrained scheduled memory and incorrect memory allocation during inference optimization. These quality issues underscore the critical importance of quality assurance for automotive DL frameworks, necessitating the urgent design of specialized automotive DL framework testing.

Unfortunately, existing DL framework testing methods \cite{muffin, lemon, gandalf} rarely succeed in detecting issues in automotive DL frameworks. Specifically, existing methods generate DL models as test inputs. However, these models lack the necessary capabilities required for deployment in autonomous driving systems and thus cannot be successfully deployed to detect bugs in automotive DL frameworks. In detail, these models lack the following three capabilities:

\textbf{(1) Multiple Input/Output Tensor Processing.} Existing methods generate models with only a single input and output tensor, limiting their ability to uncover bugs in scenarios requiring multiple input/output tensors. However, autonomous driving systems often demand such capabilities. Consequently, existing approaches fail to detect bugs arising from the complexity of multi-tensor processing.

\textbf{(2) Multi-Modal Data Processing.} The existing methods generate models limited to processing single-modality data, resulting in deployment failures when handling multi-modal data. For example, a \DIFadd{camera-LiDAR detection} model must process at least two modalities: \DIFadd{camera images and LiDAR cloud points}.
Autonomous driving systems typically require models to handle multiple data modalities. Without this capability, the models themselves may cause the system to crash, not due to issues in the automotive DL frameworks but because of limitations in the models. This results in the inability to properly test the automotive DL frameworks.

\textbf{(3) Multi-Level Data Feature Extraction.} Autonomous driving systems often require automotive DL models to output both final results and multiple data features from intermediate layers. For example, a LiDAR\DIFadd{-only} detection model is required to provide two outputs: low-level data features transmitted as intermediate results to trajectory prediction modules, and high-level data features used for LiDAR point cloud classification in perception modules.
Models that lack multi-level feature extraction capabilities can cause crashes due to issues within the model itself, hindering proper testing in automotive DL frameworks.

\DIFaddbegin \DIFadd{To equip DL models with the aforementioned capabilities, essential model structures—referred to as model components— need to be incorporated. For example, to support multiple inputs/outputs, DL models should include components with parallel substructures, known as heads. To process multi-modal data, models require components with cross-modal attention mechanisms or concatenation-based fusion layers, known as necks. To extract hierarchical features, models should include repeated feature extraction substructures, referred to as backbones. However, existing approaches typically generate DL models at the operator level rather than at the model component level, for instance, by applying mutations that alter individual operators. Such low-level mutations often disrupt essential model capabilities (e.g., multi-modal data processing), resulting in models that fail to function properly in real-world scenarios like autonomous driving. These failures often manifest as system crashes due to insufficient processing capabilities.
}

\DIFaddend To bridge the gap between underdeveloped test input model generation and multiple capabilities required by autonomous driving system, we develop Scalpel, an automotive DL framework testing tool that generates test input models at model component level capable of handling multiple input/output tensors, processing multi-modal data, and extracting features at various levels. \DIFaddbegin \DIFadd{Specifically,  
}\DIFaddend Scalpel first collects \DIFaddbegin \DIFadd{existing model components from deployed models into }\DIFaddend a repository and heuristically generates test input models through selection, mutation, and assembly. Selecting and mutating model components \DIFadd{(primarily backbones)} are guided by their contribution to bug detection. Scalpel then tests the models against an autonomous driving dataset using differential testing between cloud-end and automotive DL frameworks. Meanwhile, Scalpel identifies bugs and refines the model component repository and heuristic indicators for more effective future testing.
We compare Scalpel with three state-of-the-art methods, respectively Muffin \cite{muffin}, LEMON \cite{lemon}, and Gandalf \cite{gandalf}. The experimental results show that Scalpel surpasses existing methods in both effectiveness and efficiency. In \DIFaddbegin \DIFadd{Apollo}\DIFaddend , 16 crashes and 21 NaN \& inconsistency bugs are detected, all of which have been reported to open-source communities, with 10 crashes confirmed by developers. Root causes for detected crashes include Inference optimization, Unimplemented function, Limited memory, Incompatible argument, and Other. Detected NaN \& inconsistency bugs stem from Mismatch implementation, Precision error, and Randomness. Scalpel achieves a 27.44$\times$ improvement in model generation efficiency and an 8.5$\times$ enhancement in bug detection efficiency. \DIFaddbegin \DIFadd{In addition, Scalpel detects nine crashes and 16 NaN \& inconsistency bugs in Autoware, showing its excellent generalization.
}\DIFaddend

Our main contributions are as follows:

\begin{itemize}
\item We propose the first automotive DL framework testing method capable of generating test input models that support multiple input/output tensors, multi-modal data processing, and multi-level feature extraction.
\item \DIFaddbegin \DIFadd{We create separate model component repositories for Apollo and Autoware}\DIFaddend , offering suitable model components that are equipped with the necessary capabilities for autonomous driving systems.
\item We introduce the first model component-level test input model generation pipeline, which incorporates heuristic strategies for model component selection and mutation, along with model assembly rules.
\item We implement our method as a tool, Scalpel, to test Apollo's \DIFaddbegin \DIFadd{and Autoware's }\DIFaddend native automotive DL frameworks. Experimental results show that\DIFaddbegin \DIFadd{, in Apollo, }\DIFaddend Scalpel outperforms existing methods in both effectiveness and efficiency. Specifically, Scalpel detects 16 crashes and 21 NaN \& inconsistency bugs, while demonstrating 27.44$\times$ and 8.5$\times$ improvements in model generation efficiency and bug detection efficiency in Apollo. \DIFaddbegin \DIFadd{Additionally, Scalpel detects nine crashes and 16 NaN \& inconsistency bugs in Autoware, which shows its excellent generalization.
}\DIFaddend \end{itemize}

We release our data and code on GitHub (\url{https://github.com/DLScalpel/Scalpel}) to support reproducibility and further research.

\section{Background}
\label{background}

\subsection{DL in Autonomous Driving Systems}
\label{dl in autonomous driving system}
Deep learning (DL) has become a cornerstone of autonomous driving systems. \DIFaddbegin \DIFadd{In the DL domain, the API functions provided by DL frameworks are referred to as DL operators. These operators form the building blocks of DL model components (e.g., backbones), which in turn are assembled into complete DL models. } \DIFaddend DL models are most extensively employed in the perception module, where they enable critical tasks such as object detection \cite{objectdetection_1} and lane recognition \cite{objectdetection_2}. These perception outputs serve as the foundation for downstream rule-based decision-making modules like path planning \cite{ADSplanning} and trajectory-prediction \cite{ADStrajectory-prediction}, ensuring safe navigation by dynamically interpreting complex environments \cite{Apollo}. For example, some DL models (e.g., Smoke \cite{smoke}) process \DIFaddbegin \DIFadd{images }\DIFaddend from a single camera for object detection. Some DL models (e.g., Petr \cite{petr}) fuse \DIFaddbegin \DIFadd{camera images and LiDAR point clouds }\DIFaddend to detect pedestrians \DIFaddbegin \DIFadd{(LiDAR point clouds are typically transformed into pillars as a preprocessing step before object detection)}\DIFaddend. The seamless interaction between DL-driven perception modules and rule-based decision-making modules underscores the hybrid intelligence paradigm in autonomous driving systems, balancing data-driven adaptability with operational safety constraints \cite{dlADSremark}.

\DIFaddbegin \DIFadd{In the software development lifecycle, DL models must undergo a series of steps to function reliably in autonomous driving systems. First, models are trained using cloud-end frameworks such as TensorFlow \cite{tensorflow}, PyTorch \cite{pytorch}, and PaddlePaddle \cite{PaddleInference}. Once trained, these models are deployed to vehicle-end inference frameworks. For example, PaddleInference \cite{PaddleInference} is integrated into the Apollo, while TensorRT \cite{tensorrt} is used in Autoware.
Our approach is positioned between the training and deployment stages, serving as a validation step to detect inconsistencies between the cloud-end and vehicle-end DL frameworks.
}

\DIFaddend \subsection{Automotive DL Model Components}
\label{automotive dl model components}
Automotive DL models should be capable of: (1) processing multiple input/output tensors, (2) handling multi-modal data, and (3) extracting data features at multiple levels. To enable these functionalities, the DL model architecture introduces three key components: \textit{head}, \textit{neck} and \textit{backbone} \cite{neckdesign,backbonedesign}.

\subsubsection{Head}
\label{background_head}
Head is designed to process multiple input/output tensors. Specifically, modern autonomous driving systems rely on diverse sensors, requiring DL models to process multiple inputs (e.g., RGB images, depth maps) and generate heterogeneous outputs (e.g., bounding boxes, semantic masks). To address this, head is designed with parallelized substructures. For example, a detection head deploys separate branches for classification and regression, while a segmentation head aggregates multi-scale feature maps through skip connections. Head is usually used to dynamically adapt to varying sensor configurations and task-specific output formats, ensuring compatibility with downstream modules.

\subsubsection{Neck}
\label{background_neck}
Neck is designed to process multi-modal data. Multi-modal data processing is critical for perception module, as individual sensors exhibit complementary strengths (e.g., cameras for providing rich textural and color details, LiDAR for capturing precise geometric information). Neck bridges modality-specific feature extractors (e.g., ResNet for images \cite{resnet}, PointNet for LiDAR \cite{pointnet}) by implementing cross-modal attention mechanisms or concatenation-based fusion layers. For instance, the neck for PETR \cite{petr} aligns camera and LiDAR features in a shared latent space using transformer-based attention, enabling the model to correlate visual textures with 3D point clouds.

\subsubsection{Backbone}
\label{background_backbone}
Backbone is designed to extract data features at multiple levels. Multi-level data feature extraction is fundamental to interpreting complex driving scenes, where low-level edges, mid-level shapes, and high-level semantics collectively inform the decision-making module. Backbone achieves this ability through repeatedly stacking the same structure called block. Each block extracts data features at different abstract levels: early blocks capture local patterns (e.g., textures), while deeper blocks integrate global context (e.g., object relationships). For example, a ResNet-50 \cite{resnet} backbone iteratively applies residual blocks, downsampling feature maps while expanding receptive fields. 

 \section{Methodology}
\label{methodology}

\subsection{Overview}

\begin{figure*}[htpb]
    \centering
    \includegraphics[width=0.9\textwidth]{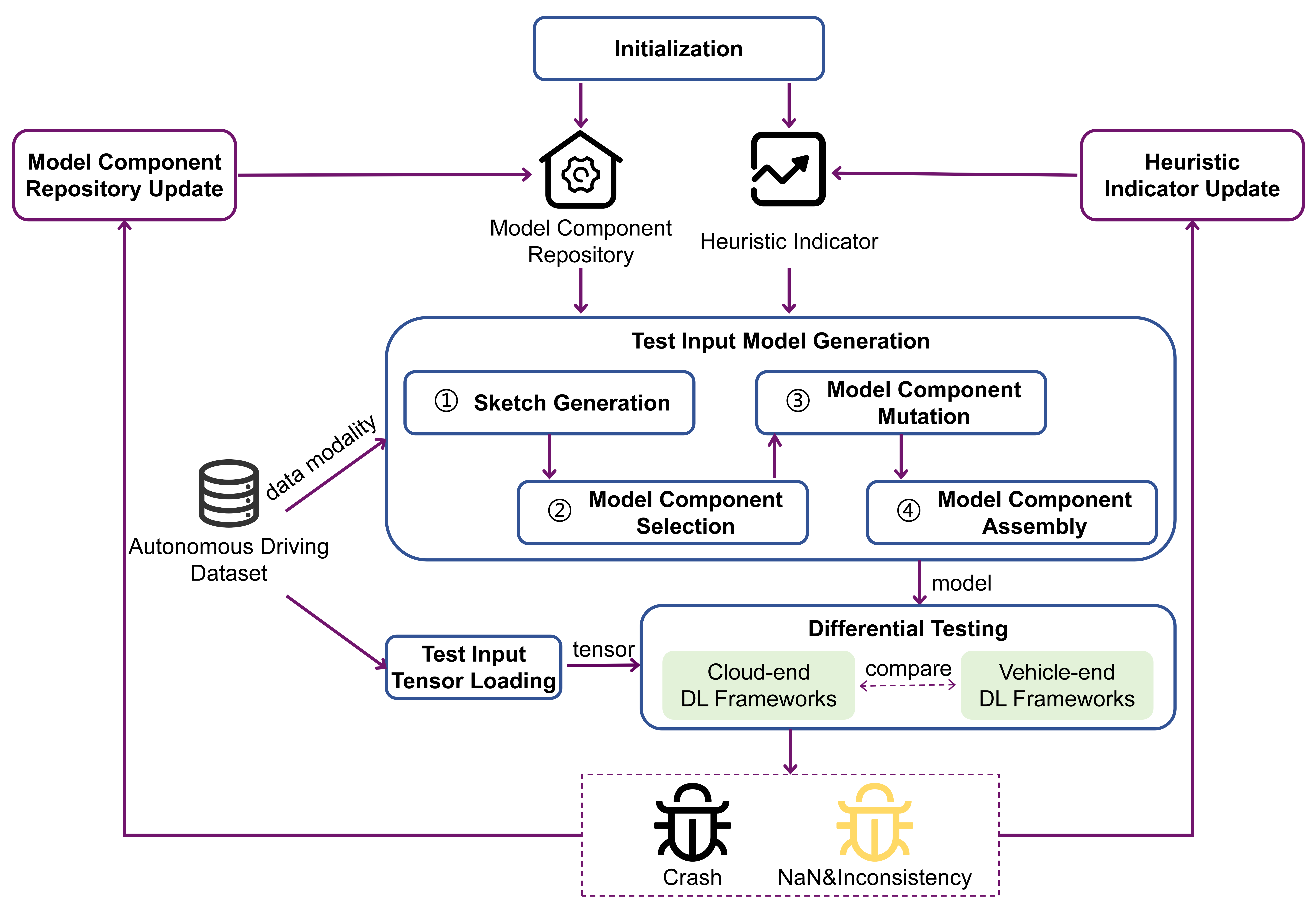}
    \Description{workflow}
    \caption{The Workflow of Scalpel
    }
    \label{workflow}
\end{figure*}

In this work, we propose Scalpel, an automotive DL framework testing method which generates test input models that are able to process multiple input/output tensors, process multi-modal data, and extract data features at multiple levels. Scalpel collects available model components in a repository, and heuristically generates new test input models by selecting, mutating, and assembling model components. Figure \ref{workflow} shows the workflow of Scalpel. In initialization stage (Section \ref{initialization}), Scalpel sets up a model component repository to maintain available model components, and initializes heuristic indicators used to enable effective model component selection and mutation.
The test input model generation consists of four steps: \ding{192} Scalpel generates model sketch based on test input data modality. Each sketch is associated with a deployment scenario and specifies the connection among model components, and it is used for model component selection and assembly (Section \ref{sketch generation}).
\ding{193} Based on the sketch, Scalpel selects model components from the model component repository, heuristically guided by each model component's 
contribution score of bug detection (Section \ref{model component selection}). 
\ding{194} \DIFdelbegin Scalpel mutates the selected model component by replacing its operator, heuristically guided by each operator's
contribution score of bug detection (Section \ref{model component mutation}).
\ding{195} Scalpel assembles mutated model components into a complete model according to the sketch (Section \ref{model component assembly}). After that, Scalpel loads \DIFaddbegin \DIFadd{test input tensors from autonomous driving datasets }\DIFaddend (Section \ref{test input tensor loading}). With generated models and loaded test input \DIFaddbegin \DIFadd{tensors}\DIFaddend , Scalpel detects automotive DL framework bugs by applying differential testing between cloud-end \DIFaddbegin \DIFadd{and vehicle-end DL frameworks (}\DIFaddend Section \ref{differential testing}). Based on detected bugs, Scalpel updates the model component repository and heuristic indicators to promote following rounds' effectiveness (Section \ref{heuristic guidance}).

\subsection{Initialization}
\label{initialization}

\subsubsection{Model Component Repository Initialization}
\label{model component repository initialization}

Scalpel generates new DL models by assembling model components. We set up a model component repository to maintain all model components available in Scalpel. In the initialization stage, we collect all models deployed in the Apollo and Autoware, and add all these model components into the model component repository. Since model components serve as the smallest functional units that preserve the model's capabilities, Scalpel categorizes components based on the capabilities they provide. Specifically, model components are divided into three types: head, neck, and backbone. The head specifies the number of input and output tensors, providing the ability to process multiple input/output tensors. The neck provides the ability to process multi-modal data. The backbone offers the ability to extract data features at multiple levels. 
In Scalpel, a model is defined as a quadruple $<sketch$, $head$, $neck$, $backbone>$. $Sketch$ is the basic skeleton of a model, describing the connection between all model components, and $head$, $neck$, and $backbone$ respectively represent three model components used to assemble a model.

By analyzing all available models in the autonomous driving system, we categorize model deployment scenarios into three types based on data modalities.
\DIFaddbegin \DIFadd{ Specifically, 
DL models in autonomous driving systems must handle two data modalities: camera images and LiDAR point clouds. Accordingly, there are three deployment scenarios: (1) Camera-only detection, where models process only camera images; (2) LiDAR-only detection, where models process only LiDAR point clouds; and (3) Camera-LiDAR detection, where models process both modalities. These scenarios are summarized in Table \ref{table_model_deployment_scenario}, with the first column listing the deployment scenarios and the second indicating the corresponding data modalities. Since processing different modalities requires different model components, Scalpel automatically maps all components to their respective deployment scenarios based on the modalities they handle. Some components, such as backbones, may be mapped to multiple scenarios due to their ability to process different modalities.
}

\begin{table}[htpb]
    \centering
    \caption{\DIFadd{Scenario Categorization by Data Modality}}
    \begin{tabular}{c|c}
    \hline
         \textbf{Scenario} & \textbf{Data Modality}\\ \hline
         Camera-only detection & camera image\\ \hline
        LiDAR-only detection & LiDAR point cloud\\ \hline
         \multirow{2}{*}{Camera-LiDAR detection} & \makecell[c]{camera image} \\ \cline{2-1}
          &LiDAR point cloud\\ \hline
    \end{tabular}
    \label{table_model_deployment_scenario}
\end{table}

\subsubsection{Heuristic Indicator Initialization}
\label{heuristic indicator initialization}
To enhance the effectiveness of test input model generation, Scalpel designs the heuristic indicator $contribution_c$ to guide the selection of model components and the heuristic indicator $contribution_{op}$ to guide the selection of operators during model component mutation. 
$Contribution_c$ describes the accumulative contribution of each component to bug detection. The contribution to bug detection refers to $\Delta eff$ in Section \ref{heuristic guidance}.
$Contribution_{op}$ describes the accumulative contribution of each operator to bug detection. In initialization, Scalpel sets all $contribution_c$ and $contribution_{op}$ to 1, which is the maximum value in the input tensor. The value of $contribution_c$ and $contribution_{op}$ will be updated after \DIFaddbegin \DIFadd{each iteration of the algorithm}\DIFaddend .

\subsection{Test Input Model Generation}
\label{test input model generation}

\DIFaddbegin \DIFadd{Scalpel's model component-based generation method is designed based on a novel observation: A model must possess specific capabilities (e.g., multi-modal data processing) to be successfully deployed in autonomous driving systems, and model components (rather than operators) constitute the minimal units preserving these capabilities. Therefore, our approach generates models by assembling real-world model components (instead of combining operators), thereby avoiding the compromise of essential model capabilities. Scalpel's detailed model generation workflow is as follows. Firstly, Scalpel generates model sketch to specify the connections between model components. Then, Scalpel selects model components from the model component repository and mutates them to generate new model components. By assembling mutated model components based on the sketch, new test input models are generated.
}

\DIFaddend \subsubsection{Sketch Generation}
\label{sketch generation}
\DIFaddbegin \DIFadd{In Section \ref{model component repository initialization}, we have summarized three model deployment scenario in the autonomous driving system. In this section, Scalpel generates models sketches for each scenario respectively. Scalpel's sketches systematically describe for each scenario: what inputs models need to process, what outputs they should produce, what essential capabilities are required to process these inputs and outputs, what model components are consequently needed to possess these capabilities, and how these model components should be assembled. Scalpel's model sketch enables automated and semantically valid mutations by enforcing constraints like tensor compatibility. This makes mutations meaningful, executable, and testable. In this section, we will introduce Scalpel's sketch template and model sketches for each scenario.
}\DIFaddend 

\begin{figure}[htpb]
    \centering
    \includegraphics[width=0.4\textwidth]{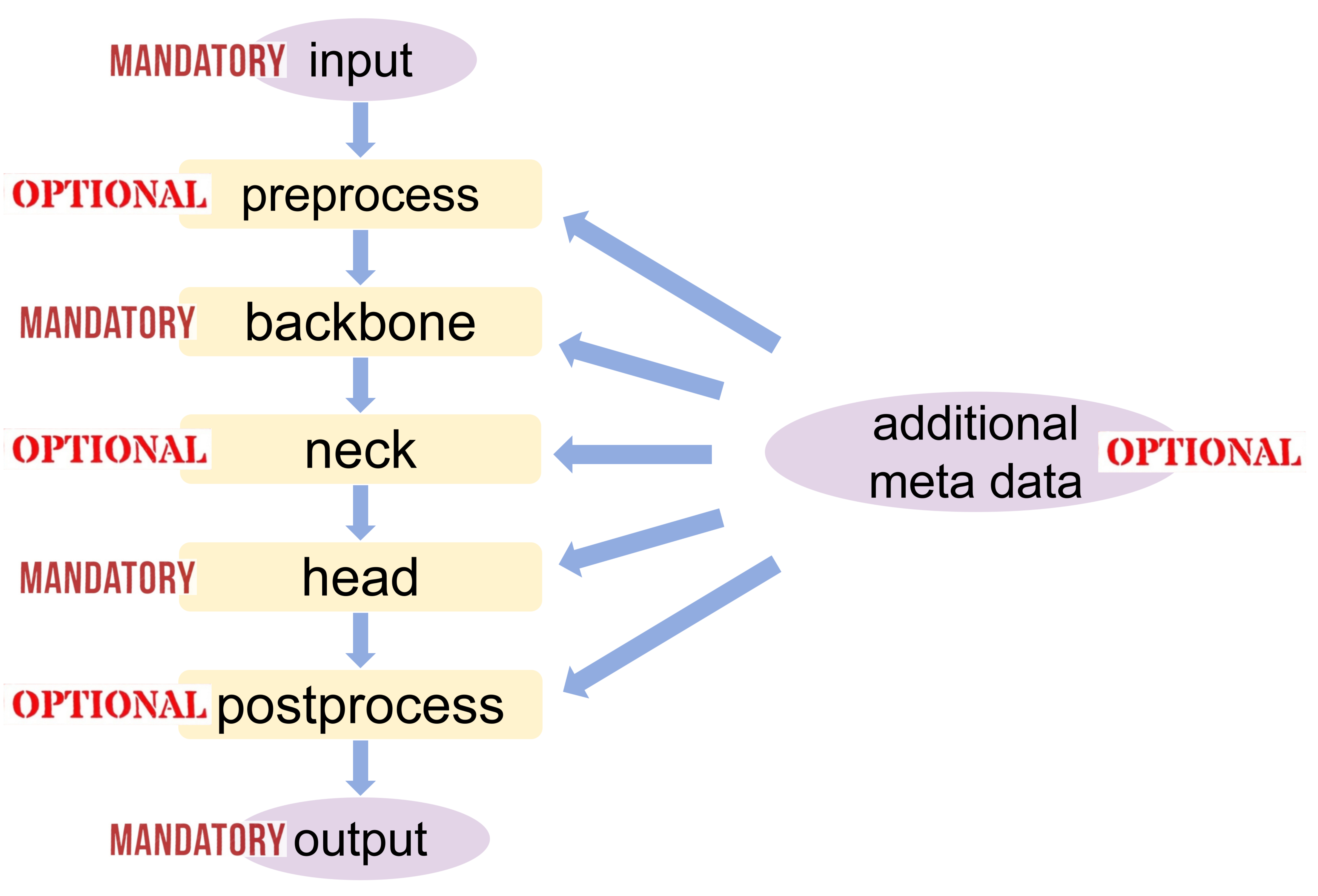}
    \Description{sketch_template}
    \caption{The Template of the Model Sketch}
    \label{sketch_template}
\end{figure}

Figure \ref{sketch_template} illustrates the sketch template in Scalpel where purple ovals denote data, while yellow rectangles denote model components. Elements with ``mandatory" indicate that the data or model components are mandatory and elements with ``optional" indicate that the data or model components are optional (optional components will be adjusted according to the model deployment scenario). The figure shows that the input and output are both mandatory data, and the additional meta data (e.g., camera configuration)
is optional. The backbone and head are mandatory model components in all scenarios. The neck is an optional model component. Additionally, in certain scenarios, it is necessary to incorporate preprocess and postprocess tailored to the data modality. For example, in \DIFaddbegin \DIFadd{LiDAR-only }\DIFaddend detection, the preprocess and postprocess are adopted to deal with LiDAR \DIFaddbegin \DIFadd{point clouds}\DIFaddend .
By adjusting the model sketch template, Scalpel generates a model sketch for each model deployment scenario.

\begin{figure}[htpb]
    \centering
    \DIFdelbeginFL 
\DIFdelendFL \DIFaddbeginFL \includegraphics[width=0.35\textwidth]{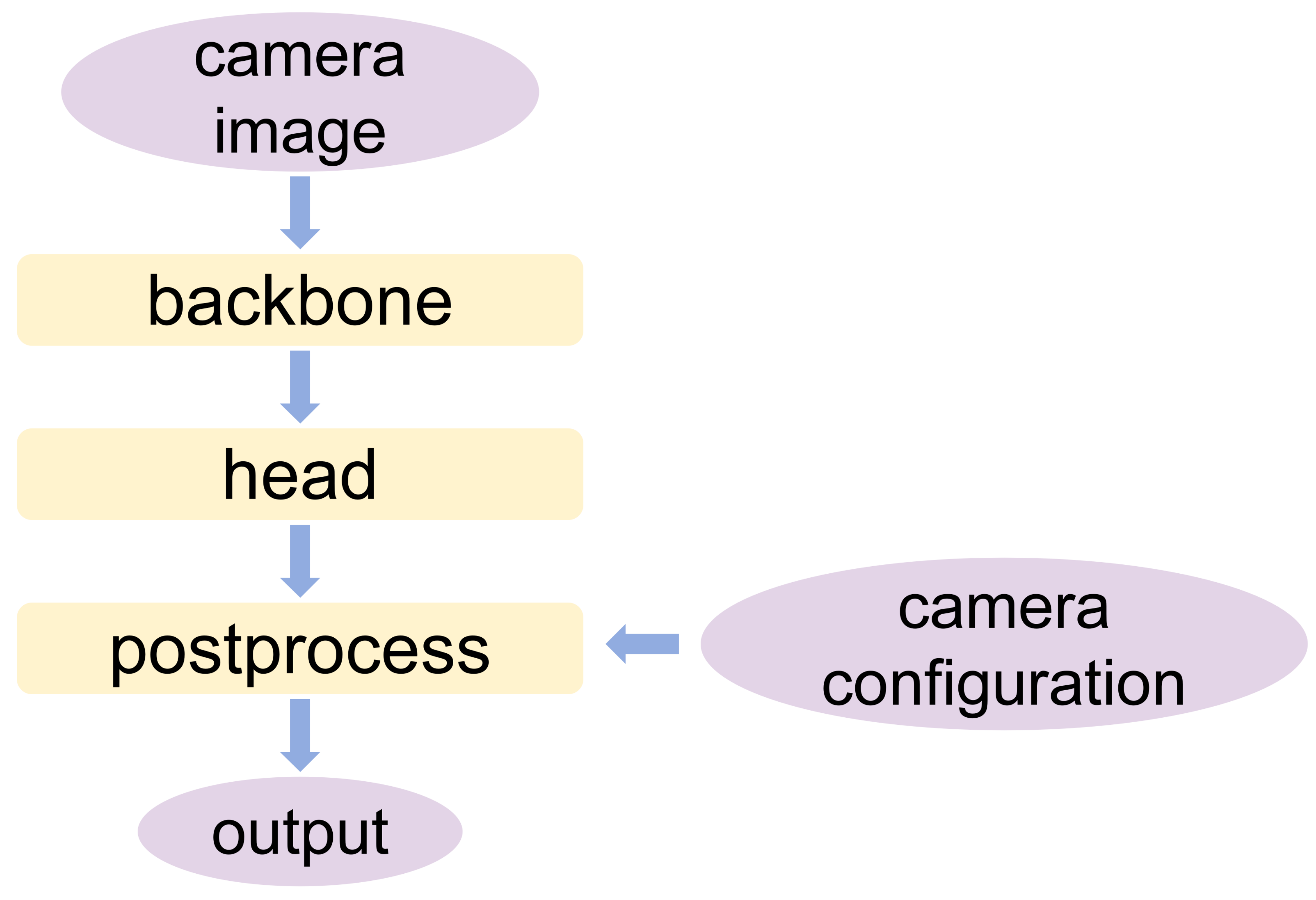}
    \DIFaddendFL \Description{mono_sketch}
    \caption{ \DIFaddbeginFL \DIFaddFL{The Sketch for Camera-only Detection}\DIFaddendFL }
    \label{mono_sketch}
\end{figure}

\DIFaddbegin \DIFadd{Figure \ref{mono_sketch} shows the model sketch for the scenario \textit{Camera-only detection}. The input for this scenario consists solely of camera images and does not require preprocessing. Since only a single data modality is processed, the neck is removed. The postprocess is retained for processing camera images based on camera configuration.
}\DIFaddend 

\begin{figure}[htpb]
    \centering
    \DIFdelbeginFL 
\DIFdelendFL \DIFaddbeginFL \includegraphics[width=0.32\textwidth]{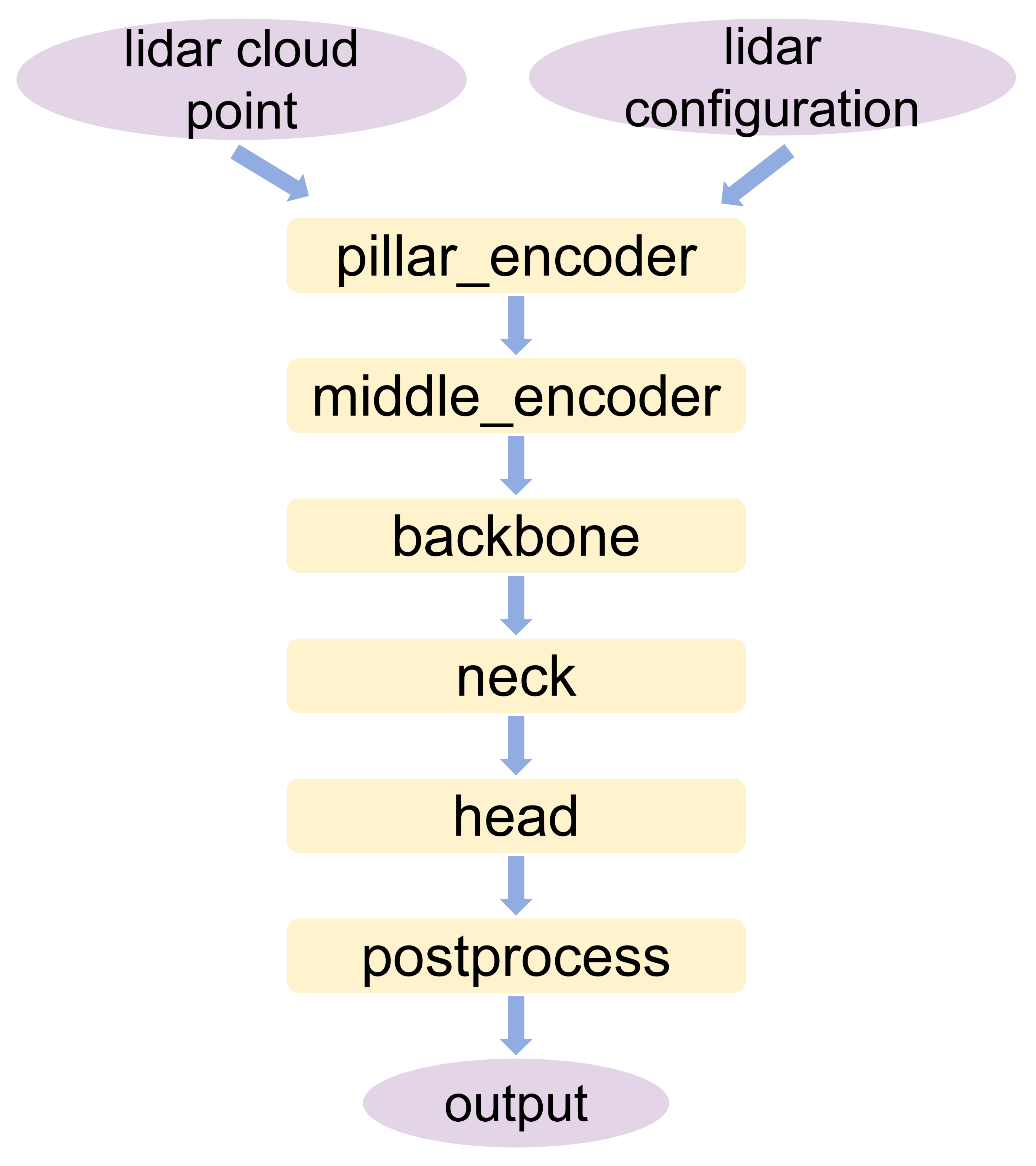}
    \Description{LiDAR_sketch}
    \DIFaddendFL \caption{\DIFaddbeginFL \DIFaddFL{The Sketch for LiDAR-only Detection}\DIFaddendFL }
    \DIFdelbeginFL 
\DIFdelendFL \DIFaddbeginFL \label{LiDAR_sketch}
\DIFaddendFL \end{figure}

\DIFaddbegin \DIFadd{Figure \ref{LiDAR_sketch} shows the model sketch for the scenario \textit{LiDAR-only detection}. The input for this scenario is LiDAR point clouds. LiDAR point clouds need preprocess (including $pillar\_encoder$ and $middle\_encoder$), which are used to divide the LiDAR points clouds into pillars. Since LiDAR point clouds are multi-modality data, the neck is retained. The postprocess is retained for processing LiDAR point clouds based on LiDAR configuration.
}

\DIFaddend 

\begin{figure}[htpb]
    \centering
    \DIFdelbeginFL 
\DIFdelendFL \DIFaddbeginFL \includegraphics[width=0.35\textwidth]{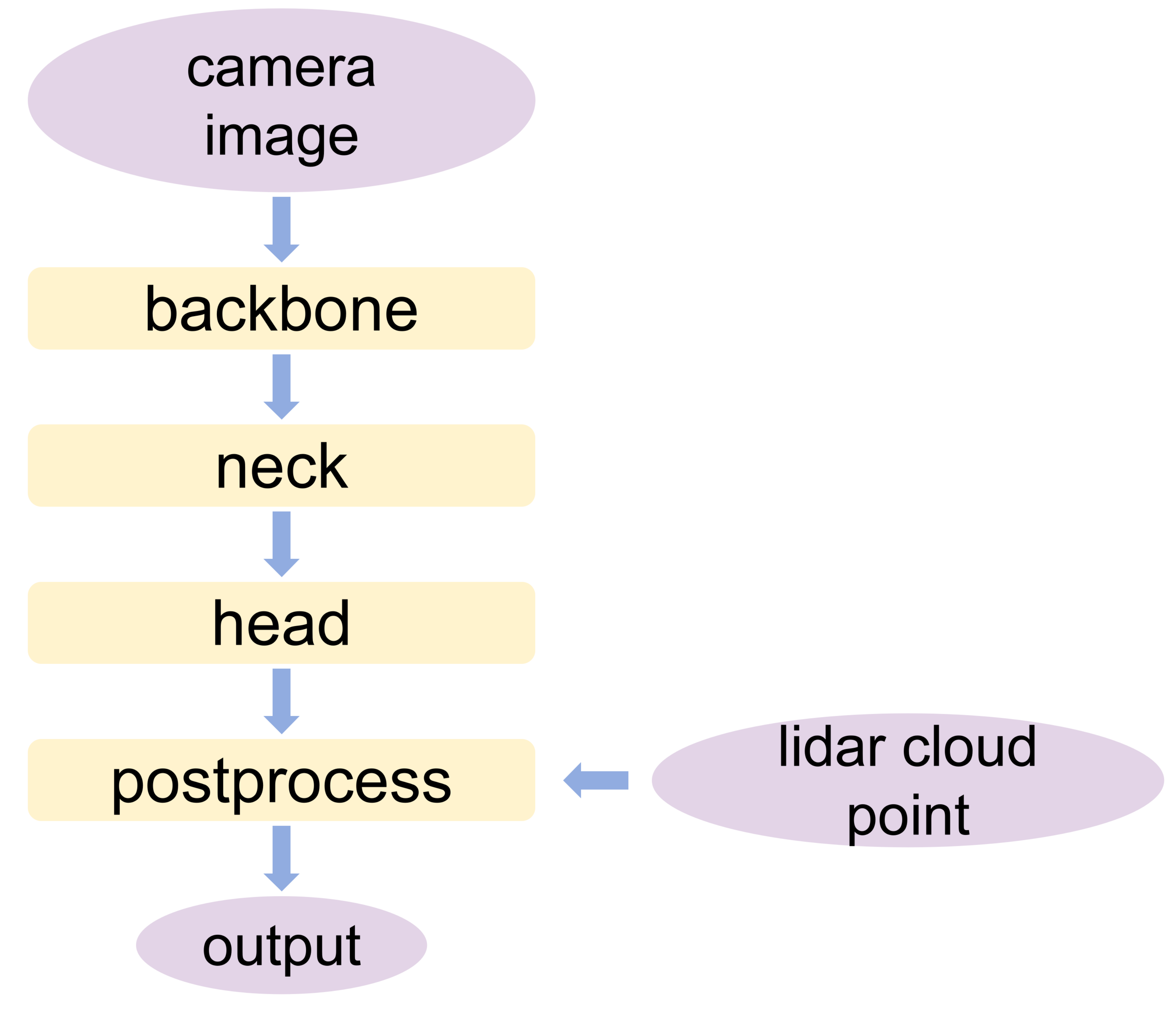}
    \Description{multi_sketch}
    \DIFaddendFL \caption{ \DIFaddbeginFL \DIFaddFL{The Sketch for Camera-LiDAR Detection}\DIFaddendFL }
 \DIFaddbeginFL \label{multi_sketch}
\DIFaddendFL \end{figure}

\DIFaddbegin \DIFadd{Figure \ref{multi_sketch} shows the model sketch for the scenario \textit{Camera-LiDAR detection}. The input for this scenario consists of camera images and LiDAR point clouds. Similar to Camera-only detection, camera images are directly processed by backbone (without preprocess). As designed in autonomous driving systems, LiDAR point clouds are simply used to aid camera images for detection in this scenario, so LiDAR point clouds are used in postprocess. Additionally, since this scenario needs to handle more than two data modalities, the neck in the template is retained.
}\DIFaddend

\subsubsection{Model Component Selection}
\label{model component selection}
Based on a given deployment scenario, Scalpel heuristically selects model components from the repository using the following four rules.

(1) Only components that can be used in the scenario of the generated sketch can be selected.

(2) Each sketch only requires selecting one head and backbone.

(3) If the model deployment scenario requires the model to process only one data modality, no neck is selected. If the model deployment scenario requires the model to process two or more data modalities, only one neck is selected.

(4) If there are multiple eligible components in the model component repository, the selection of the component is based on a weighted random process, where the probability of selecting each component is determined by its accumulative contribution to bug detection in previous rounds. Specifically, for a component $c$, its probability of being selected in the current round, $p_c$, is based on its accumulative contribution to bug detection in previous rounds:
\begin{equation}
\label{(1)}
{p}_{c}=\frac{{contribution}_{c}}{\sum ^{n}_{i=1} {{contribution}_{i}}}
\end{equation}
where n represents the total number of eligible components in the model component repository.

Note that the number of selected model components in the first two rules must strictly follow the sketch. Otherwise, the generated model will fail during deployment.

\subsubsection{Model Component Mutation}
\label{model component mutation}

Scalpel generates new model components by mutating existing ones in the model component repository. As aforementioned, model components include heads, necks and backbones. Modifying  head or neck is highly likely to produce a component that causes the assembled model to fail during deployment. By contrast, backbone is more flexible and can be mutated to generate various models for testing. Therefore, Scalpel focuses on mutating backbones.

As introduced in Section \ref{background_backbone}, automotive DL models are expected to extract data features at multiple levels. To meet this expectation, in the backbone, identical model structure often appears repeatedly, which is called block. Scalpel designs a block-based backbone mutation method to generate new backbones, as shown in Algorithm \ref{algorithm_backbone_mutation}. The input of the algorithm is the backbone selected in Section \ref{model component selection}, and the output is a set of mutated backbones. 

Scalpel begins by partitioning the backbone into groups of identical blocks to facilitate targeted mutations~(line 2). For each group, its representative block is picked and iteratively mutated~(lines 6-11) until a valid mutated block is generated. After that, Scalpel replaces all the blocks in the input backbone with the mutated block to generate a new backbone~(line 12), and the resulting backbone variant is added to the set of mutated backbones~(line 13). 

In the mutation process, each block is represented as a directed acyclic connected graph, where vertices represent tensors and edges represent operators. 
During each mutation iteration, Scalpel randomly selects a start vertex $i$ and an end vertex $j$ in the block~(lines 7-8), and heuristically selects a mutation operator $op$~(line 9). 
Then Scalpel mutates the block by replacing the operator between $i$ and $j$ with the selected mutation operator $op$~(line 10), and validates the resulting block~(line 11).
This mutation process repeats until a valid mutated block is obtained. A block is valid if and only if the block can be represented as a directed acyclic connected graph. \DIFaddbegin \DIFadd{This validation contributes to guaranteeing a model's correctness.}\DIFaddend

When selecting the mutation operator $op$, all operators available for use in a model are considered as candidates.
The probability of each operator being selected is $p_{op}$, defined as: 
\begin{equation}
\label{(2)}
     {p}_{op}=\frac{{contribution}_{op}}{\sum ^{m}_{k=1} {{contribution}_{k}}} 
\end{equation}
where $op$ represents the operator, $ {contribution}_{op}$ represents operator's accumulative contribution to bug detection in past rounds (see Section \ref{heuristic guidance}), and $m$ is the total number of available operators. If the selected operator is the same as the original operator between vertices $i$ and $j$, then Scalpel selects a new one.

\begin{algorithm}[htpb]
\caption{Backbone Mutation Algorithm}
\label{algorithm_backbone_mutation}
\begin{algorithmic}[1]
\Require $backbone$: a backbone in the component repository
\Ensure $mutatedBackbones$: a set of mutated backbones
\State $mutatedBackbones \leftarrow \emptyset$
\State $blockGroups \leftarrow groupBlocks(backbone)$ \newline \Comment{Partition the input backbone into groups of identical blocks}
\ForAll {$group \in blockGroups$} \Comment{Mutate blocks}
    \State $block \leftarrow getBlock(group)$
    \State $block' \gets null$
    \Repeat
        \State $i \leftarrow selectStartVertex(block)$
        \State $j \leftarrow selectEndVertex(block)$
        \State $op \leftarrow selectOperatorForMutation()$ 
        \State $block' \leftarrow mutateOperator(block,i,j,op)$
    \Until{$isValid(block') == true$}
    \State $backbone' \leftarrow replaceBlock(backbone,group,block, block')$
    \State $mutatedBackbones \leftarrow mutatedBackbones \cup backbone'$
\EndFor
\State \Return $mutatedBackbones$
\end{algorithmic}
\end{algorithm}

\subsubsection{Model Component Assembly}
\label{model component assembly}

Using the selected head and neck components (Section \ref{model component selection}) along with the mutated backbones (Section ~\ref{model component mutation}), Scalpel assembles them according to the generated sketch (Section \ref{sketch generation}) \DIFaddbegin \DIFadd{to generate }\DIFaddend input models for testing. 
Multiple backbones are generated, with each backbone being used once to assemble a model.
As the neck is optional, there are two different assembly cases.
For each case, the parameter settings between model components must follow specific rules.

When neck is required, the assembly needs to follow the three rules listed below: \ding{192} The channel dimensions of the neck's output tensor(s) must match those of the head's input tensor(s). \ding{193} The channel dimensions of the backbone's output tensor(s) must match the neck's input tensor(s). \ding{194} If the backbone generates multiple output tensors, the number of input tensors to the neck must match this quantity.

When neck is excluded, the two rules below must be followed.
\ding{192} The channel dimensions of the backbone's output tensor(s) must match those of the head's input tensor(s).
\ding{193} If the backbone generates multiple output tensors, the number of input tensors to the head must match this quantity.

\subsection{Test Input Tensor Loading}
\label{test input tensor loading}

In Scalpel, the test input tensor is loaded from the autonomous driving dataset (e.g., KITTI \cite{KITTI}). Before loading the tensor, \DIFaddbegin \DIFadd{Scalpel }\DIFaddend preprocesses the meta data in the dataset to meet the requirements of automotive DL models. Existing framework testing methods load only one test input tensor (containing a single data modality) for each DL model. In contrast, an automotive model often require multiple test input tensors (usually containing multiple data modalities). \DIFaddbegin \DIFadd{The preprocessing varies for different data modalities, following the existing practices in Apollo and Autoware. Specifically, for camera images, Scalpel scales and crops the images according to the dimensions required by the autonomous driving system. For LiDAR point clouds, Scalpel converts them to voxels, which contributes to better extracting data features \cite{voxelization}.
}\DIFaddend 

\subsection{Differential Testing}
\label{differential testing}
\DIFaddbegin \DIFadd{With generated input models (Section \ref{test input model generation}) and loaded input tensors (Section \ref{test input tensor loading}),
Scalpel performs differential testing between cloud-end and vehicle-end frameworks to detect bugs. Specifically, Scalpel firstly executes models on cloud-end DL frameworks and removes models that crash. In this way, all remaining models can be successfully executed on the cloud end, which guarantees the correctness of models executing on the vehicle. Unlike models deployed on the cloud, models deployed on the vehicle often output multiple tensors \cite{Apollo}, rendering the test oracle of existing framework testing methods inapplicable. To address this, we design our new test oracles. Following existing works \cite{muffin,gandalf}, Scalpel focuses on two types of bugs: crashes and NaN \& inconsistency bugs. 
}\DIFaddend 

\DIFaddbegin \DIFadd{\subsubsection{Crash Detection}
\label{crash_detection}
Crashes are detected by analyzing execution logs. Specifically, when a failure occurs, Scalpel automatically saves the execution log. In log analysis, Scalpel first uses automated API name matching to prevent false positives, and employs automated clustering to remove redundant crash logs. For the remaining crash logs, Scalpel first applies automated bug pattern matching to preliminarily identify the root causes of bugs, and subsequently confirms the root causes of bugs through manual review. The specific process is as follows.
}\DIFaddend

\DIFadd{Scalpel first applies automatic API name matching to prevent false positives. It matches tested framework API names with stack traces in crash logs. Crashes without these API names are discarded as non-framework bugs. Next, Scalpel uses automatic clustering with text cosine similarity \cite{text_similarity} as the distance metric. Cosine similarity ranges from [-1,1], where 1 means identical logs, -1 means opposite logs, and 0 means unrelated logs. Crash logs with similarity more than 0.9 are grouped together. Only one log per group is kept to avoid redundancy. We have checked the clustering results and found that no redundant crashes are split across groups, and no distinct crashes are merged into the same group, which proves Scalpel's clustering works well. Then, Scalpel applies automatic bug pattern matching \cite{bug_pattern} to find preliminary root causes. It matches bug patterns with specific keywords in crash logs. For example, when logs contain ``CUDNN\_STATUS\_NOT\_SUPPORTED", Scalpel classifies it as a CUDNN error. Based on above preliminary bug patterns, two researchers independently refer to framework API documents and open-source community discussions to finally identify each bug's root cause. If they disagree on bug's root cause, community developers help reach final consensus.}

\subsubsection{\DIFadd{NaN \& Inconsistency Bug Detection}}
\DIFaddend Scalpel identifies NaN bugs when the output tensors on the vehicle contain NaN while those in the cloud do not, or vice versa.
Inconsistency bugs are identified by measuring the discrepancy between cloud-end and vehicle-end DL frameworks. Firstly, Scalpel groups output tensors from all frameworks based on their labels, ensuring that tensors with the same label are categorized together. 
Within each group, the output tensor from each framework is recorded. Scalpel then computes the differences between these tensors within the group. For each group, Scalpel calculates the maximum scalar value of these differences. Finally, Scalpel identifies the highest inconsistency value across all groups, referred to as $max\_inconsistency\_value$.

\DIFadd{Scalpel }\DIFaddend detects an inconsistency bug when the highest inconsistency value $max\_inconsistency\_value$ exceeds the pre-specified threshold $\epsilon$.  \DIFaddbegin \DIFadd{Following prior works (e.g., Predoo [49], Gandalf [30]), Scalpel sets $\epsilon$ to the maximum threshold used to determine approximate tensor equivalence. According to Apollo's and Autoware's official implementation, this value is 0.1. }\DIFaddend In this manner, all inconsistencies detected by Scalpel are assured to be identified as different tensors within the autonomous driving system.

\subsection{Heuristic Guidance}
\label{heuristic guidance}
Heuristic guidance is designed to improve the effectiveness of test input model generation in subsequent rounds. Scalpel updates the heuristic indicators to guide the selection and mutation of model components in future rounds. Additionally, Scalpel updates the model component repository to expand the available components for future testing.

\subsubsection{Heuristic Indicator Update}
\label{heuristic indicator update}
Scalpel defines two distinct contribution metrics: $contribution_c$ for model component selection and $contribution_{op}$ for model operator mutation. Both metrics are updated based on the effectiveness of the model in detecting bugs during the current round. In this paper, we use \textit{eff} to denote the effectiveness, which measures the model's ability to detect bugs. For crashes and NaN bugs, the effectiveness value is 0 when no bugs are triggered, and it is the mean scalar value within the test input tensor when crashes or NaN bugs occur.
For inconsistency bugs, the value of \textit{eff} is $max\_inconsistency\_value$ defined in Section \ref{differential testing}.

Based on the effectiveness, $contribution_c$ of the selected model component is updated as follows:
\begin{equation}
\label{(3)}
     {contribution}^{'}_{c}={contribution}_{c}+ \Delta eff 
\end{equation}
where ${contribution}^{'}_{c}$ and $contribution_c$ respectively represent the updated and previous accumulative contribution of the selected model component. $\Delta eff$ is defined as $eff_{new}-eff_{old}$ where $eff_{new}$ and $eff_{old}$ respectively denote the effectiveness of the newly assembled model in this round and the model assembled by components before mutation.

Similarly, based on the effectiveness, $contribution_{op}$ of the selected operator during component mutation is updated as follows:
\begin{equation}
\label{(4)}
{contribution}^{'}_{op}={contribution}_{op}+ \Delta eff
\end{equation}
where ${contribution}^{'}_{op}$ and $contribution_{op}$ respectively represent the updated and previous accumulative contributions of the selected operator. The definition of $\Delta eff$ follows equation (3).

\subsubsection{Model Component Repository Update}
\label{model component repository update}
To expand the model component repository, Scalpel adds some mutated components from successfully assembled models. Specifically, when a model assembled in a round detects a DL framework bug, the mutated component is added to the repository. If no bug is detected, components are only added if the $\Delta eff$ (defined in Section \ref{heuristic indicator update}) is greater than 0. Otherwise, the mutated components are not included. \section{Evaluation}
\label{evaluation}

In this section, we evaluate the effectiveness and efficiency of Scalpel. We compare Scalpel with three state-of-the-art methods, LEMON \cite{lemon}, Muffin \cite{muffin}, and Gandalf \cite{gandalf}. \DIFaddbegin \DIFadd{These baselines are testing methods for cloud-end DL frameworks (e.g., TensorFlow, PyTorch). Since no automotive-specific DL framework testing methods exist, these baselines are the most relevant for Scalpel. We do not include DeepMutation \cite{deepmutation} as it is significantly less effective than selected baselines. Additionally, NAS (Neural Architecture Search) focuses on generating high-performing, functionally valid models by optimization for accuracy or efficiency rather than exposuring bugs. Therefore, it rarely triggers DL framework bugs and is not considered as our baseline.
}\DIFaddend 

\subsection{Experimental Setup}
\label{experimental setup}

We download the source code of all baselines from open-source repositories and execute the code on the same workstation. When executing the source code, the parameter settings are kept consistent with those introduced in their papers. The workstation runs on Ubuntu 18.04, equipped with an RTX 4090D GPU, an Intel(R) Core(TM) i7-12700KF (12 cores, 2.4GHz) CPU. The workstation's CUDA version is 12.4, CUDNN version is 9.6.0, and NVIDIA-driver version is 535.216.01. During execution, we save the models generated by all methods. \DIFaddbegin \DIFadd{After that, we deploy these models into Apollo v9.0 for differential testing. The test input tensors for differential testing are loaded from the KITTI \cite{KITTI} and Nuscenes \cite{nuscenes} dataset. The deployed frameworks include: PaddlePaddle v2.6.2, PaddleInference v2.6.2. In addition, to evaluate Scalpel's generalization to other ROS-based autonomous driving systems, we also deploy generated models into Autoware. The deployed automotive DL framework in Autoware is TensorRT v8.5.3.1.
}\DIFaddend

\subsection{Research Questions}
\label{research questions}

In this paper, we evaluate Scalpel to answer the following four research questions.
\begin{itemize}
\item \textbf{RQ1:} How effective is Scalpel in detecting automotive DL framework bugs?

\item \textbf{RQ2:} What are the root causes of the detected automotive DL framework bugs?

\item \textbf{RQ3:} How does Scalpel perform in terms of efficiency and test coverage?

\item \textbf{RQ4:} To what extent do heuristic model component selection and mutation contribute to improving the testing effectiveness of the framework?
\end{itemize}

\subsection{RQ1: Bug Detection Effectiveness}
\label{bug detection effectiveness}

Following existing methods \cite{lemon, muffin}, Scalpel detects crashes and NaN \& inconsistency bugs in automotive DL frameworks. Among them, crashes are identified by analyzing execution logs. NaN \& inconsistency bugs are identified by comparing output tensors between cloud-end and automotive DL frameworks. Specifically, an NaN bug is detected if NaN appears in the automotive DL framework but not in the cloud-end DL framework, or vice versa. An inconsistency bug is detected if the maximum difference exceeds the threshold $\epsilon$ (Section \ref{differential testing}). \DIFaddbegin \DIFadd{When an NaN \& inconsistency bug is triggered, Scalpel automatically records the model that triggers the bug. Bugs triggered by identical models are regarded as the same bug. Models with different configurations are treated as distinct models. A single model triggers one framework bug at most.
}\DIFaddend 

We execute Scalpel and state-of-the-art methods on the same workstation for six hours. Table \ref{effectiveness_table_baseline} shows the number of unique bugs detected by Scalpel and existing methods. The first column of the table indicates the method name, while the second and third columns represent the number of crashes and NaN \& Inconsistency bugs detected, respectively. As shown in the table, Scalpel successfully detects 16 crashes and 22 NaN \& inconsistency bugs. All detected bugs have been reported to the open-source community, with 10 crashes confirmed by developers and the rest still awaiting confirmation. Muffin detects only two crashes and two NaN \& Inconsistency bugs, Gandalf detects only three crashes, and LEMON does not detect any bug. \DIFaddbegin \DIFadd{All bugs identified by baselines are detected by Scalpel, which shows Scalpel's outstanding effectiveness in bug detection.} \DIFaddend 

\begin{table}[htpb]
    \caption{Comparison in Bug Detection Effectiveness}
    \centering
    \begin{tabular}{ccc}
    \hline
         \textbf{Method} & \textbf{Crash} & \textbf{NaN \& Inconsistency}\\ \hline
         Scalpel & 16 & 22\\ 
         LEMON & 0 & 0\\ 
         Muffin & 2 & 2\\ 
         Gandalf & 3 & 0\\ \hline
    \end{tabular}
    \label{effectiveness_table_baseline}
\end{table}

To better understand the reasons behind Scalpel's superior bug detection effectiveness, we analyze the number of valid models generated by Scalpel and all baselines within a six-hour period, and the results are shown in Table \ref{model generation_table_baseline}. A valid model is one that executes successfully in the automotive DL framework, without crashing due to failing to meet the requirements of the autonomous driving system. The first column of the table lists the method names, the second column shows the total number of models generated, the third column indicates the number of valid models, and the last column represents the proportion of valid models to the total models. The experimental results show that Scalpel's rate of generating valid models is 87.07\%, significantly higher than that of LEMON (0\%), Muffin (2.65\%), and Gandalf (3.09\%). Scalpel successfully generates the largest number of valid test input models (512), which contributes to its success in bug detection effectiveness. The majority (over 96\%) of models generated by existing methods are invalid because they do not consider the requirements of the autonomous driving system during test input model generation. For instance, these models can only handle a single input or output, causing failure in the camera-LiDAR detection scenario, or they lack the ability to process multi-modal data, leading to failure when handling LiDAR point clouds.

\begin{table}[htpb]
    \caption{Comparison in Model Generation}
    \centering
    \begin{tabular}{cccc}
    \hline
         \textbf{Method} & \textbf{Total Model} & \textbf{Valid Model} & \textbf{Valid Model Rate}\\ \hline
         Scalpel & 588 & 512 & 87.07\%\\ 
         LEMON & 338 & 0 & 0\%\\ 
         Muffin & 452 & 12 & 2.65\%\\ 
         Gandalf & 583 & 18 & 3.09\%\\ \hline
    \end{tabular}
    \label{model generation_table_baseline}
\end{table}

\begin{center}
\fcolorbox{black}{lightgray}{\parbox{.95\linewidth}{\textit{Answer to RQ1:} Scalpel successfully detects 16 crashes and 21 NaN \& inconsistency bugs in the automotive DL framework, significantly surpassing existing methods. The success of Scalpel in bug detection is attributed to its higher rate of generating valid models, which can execute successfully in more model deployment scenarios and detects DL framework bugs.}}
\end{center}

\subsection{RQ2: Root Causes}
\label{root cause}

\subsubsection{Root Causes of Crashes}
\label{root cause of crashes}
\begin{figure}[htpb]
    \centering
    \includegraphics[width=0.45\textwidth]{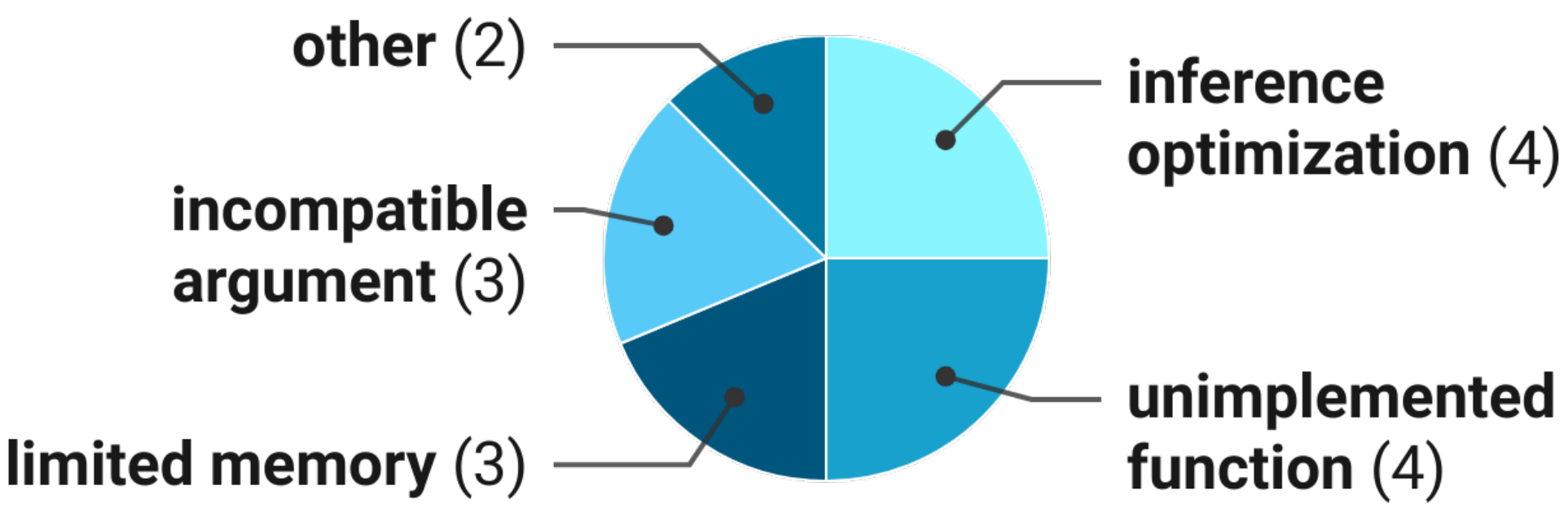}
    \caption{Root Cause of Detected Crashes}
    \label{figure_root_cause_crashes}
\end{figure}

As illustrated in Figure \ref{figure_root_cause_crashes}, Scalpel identifies five primary root causes of crashes: inference optimization, unimplemented function, limited memory, incompatible arguments, and other factors.

\textbf{Inference optimization (4/16, 25\%).} These bugs stem from defects in the inference optimization mechanisms of automotive DL frameworks. Among them, three bugs originate from memory allocation failure. Specifically, tensors are not loaded into memory correctly, which results in tensor dimension mismatch. Another bug arises from memory optimization mistake. Specifically, data that should have participated in following computation is mistakenly identified as useless data and discarded.

\textbf{Unimplemented function (4/16, 25\%).} These bugs arise from functions unsupported by automotive DL frameworks. Specifically, compared to frameworks like PaddlePaddle, automotive DL frameworks are more lightweight, and certain functions are either unsupported or only partially supported. Crashes occur when these functions are invoked. For example, PaddleInference does not support constructing a memory descriptor using a format tag.

\textbf{Limited memory (3/16, 18.75\%).} These bugs arise from limited scheduled memory. Memory exhaustion originates from unstable deployment environment in autonomous driving systems, resulting in the framework failing to fully utilize computational resources.

\textbf{Incompatible argument (3/16, 18.75\%).} These bugs stem from incompatible parameter configuration between PaddlePaddle and PaddleInference. Specifically, when DL models are converted from PaddlePaddle to PaddleInference in autonomous driving systems, some parameters within the models are not properly configured. For example, PaddleInference does not support the parameter called $shape\_range\_info\_path$.

\textbf{Other (2/16, 12.5\%).} These bugs arise from some other root causes. One crash is triggered when there is a version mismatch between NVIDIA-driver, CUDNN and CUDA. The other crash is triggered when there is a version mismatch between CUDA and autonomous driving system. These versions are declared as supported in the official documentation, but have caused crashes.

\subsubsection{Root Cause of NaN \& Inconsistency Bugs}
\label{root cause of nan and inconsistency bugs}

\begin{figure}[htpb]
    \centering
    \includegraphics[width=0.45\textwidth]{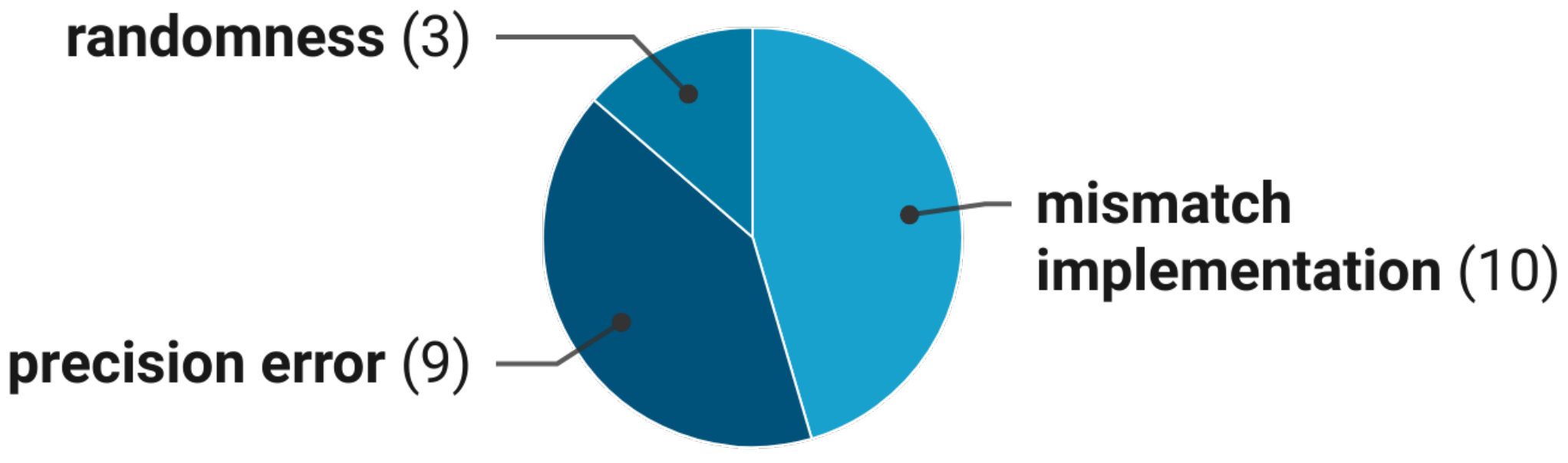}
    \caption{Root Cause of Detected NAN\&Inconsistency Bugs}
    \label{figure_root_cause_inconsistency}
\end{figure}

As shown in Figure \ref{figure_root_cause_inconsistency}, the NaN \& inconsistency bugs detected by Scalpel primarily originate from three categories of root causes: mismatch implementation, precision error, and randomness.

\textbf{Mismatch implementation (10/21, 47.62\%).} Due to code pruning, the automotive DL framework does not support some parameter settings commonly used in other DL frameworks. When converting models from other frameworks to the automotive DL framework, parameter mismatches occur, leading to NaN and inconsistency bugs. For example, PaddleInference's unique padding mode in the operator $Convolution$, and unique memory allocation in the operator $BatchNorm$.

\textbf{Precision error (9/21, 42.86\%).} Even for the same data type (e.g., bfloat16), PaddleInference employs a unique data representation compared to other DL frameworks. Additionally, defects exist in PaddleInference's quantization mechanism. The above issues lead to precision error, thereby causing inconsistency bugs.

\textbf{Randomness (3/21, 14.29\%).} In addition to the aforementioned root causes, three inconsistency bugs are triggered by random data fluctuations, which may be related to the instability of the autonomous driving system's deployment environment.

Based on detected bugs, three suggestions are proposed for antomotive DL framework developers and community. (1) Strengthen cross-framework compatibility testing for parameter configuration and operator implementation to resolve mismatches during model conversion (e.g., padding modes, memory allocation in Convolution/BatchNorm). (2) Standardize data representations and refine quantization mechanisms to align with mainstream frameworks, reducing precision error caused by unique data representation or flawed quantization mechanism. (3) Implement version dependency checks and environment stability safeguards to mitigate crashes triggered by mismatched driver/CUDA versions or deployment environment fluctuations.

\begin{center}
\fcolorbox{black}{lightgray}{\parbox{.95\linewidth}{\textit{Answer to RQ2:} In Scalpel, the root causes of detected crashes include: inference optimization, unimplemented function, limited memory, incompatible argument, and other. The root causes of detected NaN \& inconsistency bugs include: mismatch implementation, precision error, and randomness.}}
\end{center}

\subsection{RQ3: Efficiency \& Test Coverage}
\label{efficiency}

In this section, we evaluate Scalpel's efficiency, as summarized in Table \ref{table_efficiency}. The first column lists the method names, while the second and third columns present the average time required to generate a valid test input model and to detect an automotive DL framework bug, respectively. Since LEMON fails to generate valid test input models or detect any bugs, its results are marked as N/A. Experimental results show that Scalpel outperforms all existing methods, generating valid test input models in 42.19s and detecting bugs in 568.42s on average. It improves model generation efficiency by 27.44 times and bug detection by 8.5 times. This efficiency boost stems from its higher valid model rate (Table \ref{model generation_table_baseline}).

\begin{table}[htpb]
    \caption{Comparison in Efficiency}
    \centering
    \begin{tabular}{ccc}
    \hline
         \textbf{Method} & \textbf{Time Per Model (s)} & \textbf{Time Per Bug (s)} \\ \hline
         Scalpel & 42.19 & 568.42 \\ 
         LEMON & N/A & N/A \\ 
         Muffin & 1,800 & 5,400 \\ 
         Gandalf & 1,200 & 7,200 \\ \hline
    \end{tabular}
    \label{table_efficiency}
\end{table}

We calculate the test coverage to evaluate Scalpel's test sufficiency. Unlike traditional software, DL is data-driven rather than control logic-driven. As a result, traditional test coverage metrics (e.g., code coverage, branch coverage) are not suitable for measuring DL framework testing. Following existing methods \cite{comet}, we adopt operator coverage (also called API coverage) to evaluate the test sufficiency of Scalpel. Operator coverage represents the rate of operators covered by generated valid models relative to all available operators. The higher the operator coverage, the more sufficient the testing. Additionally, as a component-level model generation method, Scalpel also calculates model component coverage. Model component coverage represents the proportion of model components covered by generated valid models relative to the total number of model components in the model component repository. The higher the model component coverage, the more sufficient the testing. The experimental results are shown in Table \ref{table_coverage}. The first column in the table indicates the method name, the second column represents operator coverage, and the third column represents model component coverage. The experimental results demonstrate that Scalpel's operator coverage (100\%) and model component coverage (82\%) are both higher than all baselines, indicating that Scalpel's testing is more sufficient. In addition, LEMON, Muffin, and Gandalf do not successfully cover any model component, which shows the limitation of operator-level model generation.

\begin{table}[htpb]
    \caption{Comparison in Test Coverage}
    \centering
    \begin{tabular}{ccc}
    \hline
         \textbf{Method} & \textbf{Operator Coverage} & \textbf{Component Coverage} \\ \hline
         Scalpel & 100\% & 82\% \\ 
         LEMON & 0\% & 0\% \\ 
         Muffin & 60.71\% & 0\% \\ 
         Gandalf & 35.71\% & 0\% \\ \hline
    \end{tabular}
    \label{table_coverage}
\end{table}

\begin{center}
\fcolorbox{black}{lightgray}{\parbox{.95\linewidth}{\textit{Answer to RQ3:} Scalpel outperforms all the existing methods with the shortest average time to generate each valid test input model (42.19 seconds) and detect each automotive DL framework bug (568.42 seconds). Furthermore, it achieves the highest operator coverage (100\%) and model component coverage (82\%).}}
\end{center}

\subsection{RQ4: Ablation Study}
\label{ablation study}
Scalpel heuristically selects and mutates model component (Section \ref{model component selection} and \ref{model component mutation}). In this section, we set up three baselines to evaluate the contribution of heuristic guidance. Among them, the first is $Scalpel_c$, which randomly selects operator during mutation and only heuristically selects the model component. The second is $Scalpel_o$, which randomly selects the model component and only mutates the model component heuristically. The last is $Random$, which disables all heuristic guidance, and selects and mutates model components randomly. All baselines are executed on the same workstation for six hours separately. The experimental result is shown in Table \ref{table_ablation study}.  The first column in the table represents the method names, while the second and third columns indicate the number of crashes and NaN \& inconsistency bugs detected, respectively. The experimental results show that $Scalpel_c$ detects seven crashes and 16 NaN \& inconsistency bugs, outperforming $Random$, which highlights the importance of heuristic model component selection. $Scalpel_o$ detects eight crashes and 13 NaN \& Inconsistency bugs, surpassing $Random$, demonstrating the contribution of heuristic model component mutation. $Scalpel$ outperforms both $Scalpel_c$ and $Scalpel_o$, showing that heuristic guidance in model component selection and mutation can work together effectively.

\begin{table}[htpb]

    \caption{Bug Detection Effectiveness in Ablation Study}
    \centering
    \begin{tabular}{ccc}
    \hline
         \textbf{Method} & \textbf{Crash} & \textbf{NaN \& Inconsistency} \\ \hline
         $Scalpel$ & 16 & 22 \\ 
         $Scalpel_c$ & 7 & 16 \\ 
         $Scalpel_o$ & 8 & 13 \\ 
         $Random$ & 5 & 2 \\ \hline
    \end{tabular}
    \label{table_ablation study}
\end{table}
\begin{center}
\fcolorbox{black}{lightgray}{\parbox{.95\linewidth}{\textit{Answer to RQ4:} $Scalpel_c$ surpasses $Random$, emphasizing the importance of heuristic model component selection. Similarly, $Scalpel_o$ outperforms $Random$, demonstrating the effectiveness of heuristic model component mutation. Moreover, $Scalpel$ outperforms both $Scalpel_c$ and $Scalpel_o$, indicating that the two heuristic strategies complement each other synergistically.
}}
\end{center}

\DIFdelend \DIFaddbegin \DIFadd{\subsection{Discussion: Generalization to Autoware}
To evaluate Scalpel's generalization to other autonomous driving systems, we extend Scalpel to test the most widely used ROS-based autonomous driving systems, Autoware. The experimental settings are introduced in Section \ref{experimental setup}. Since our baselines do not support Autoware's TensorRT, we only evaluate our method's effectiveness. Table \ref{table_evaluation_discussion_autoware} shows Scalpel's performance on Autoware. In this table, the first two columns respectively represent the crash number (\#Crash) and NaN \& inconsistency number (\#N\&I) detected by Scalpel in Autoware's TensorRT. The third column represents Valid Model Rate (VMR), which means the proportion of models that can execute successfully when deployed in Autoware. The fourth column represents Time Per Model (TPM), which means the time consumption for Scalpel to generate each distinct model on Autoware. The fifth column represents Time Per Bug (TPB), which means the time consumption for Scalpel to detect each distinct bug on Autoware. The last two columns respectively represent Scalpel's Operator Coverage (OC) and Component Coverage (CC) on Autoware.
}\DIFaddend

\DIFaddbegin \begin{table}[htpb]

    \caption{\DIFaddFL{Scalpel's Performance on Autoware}}
    \centering
    \begin{tabular}{cc|c|cc|cc}
    \hline
         \textbf{\#Crash} & \textbf{\#N\&I} & \textbf{VMR} & \textbf{TPM (s)} & \textbf{TPB (s)} & \textbf{OC} & \textbf{CC} \\ \hline
         9 & 16 & 100\% & 16.63 & 864 & 100\% & 80\%\\ \hline
    \end{tabular}
    \label{table_evaluation_discussion_autoware}
\end{table}

\DIFadd{As shown in this table, Scalpel detects nine crashes and 16 NaN \& inconsistency bugs in Autoware's TensorRT. All detected crashes have been reported to open-source communities. Scalpel analyzes the crash logs as described in Section \ref{crash_detection}. Among them, five crashes originate from limited memory. As discussed in Section \ref{root cause}, due to the unstable deployment environment in Autoware, computational resources could not be fully utilized, leading to resource exhaustion. Two crashes stem from unimplemented functions, with one originating from the invocation of an undefined $Convolution$ layer in TensorRT, and the other from the invocation of an undefined $Node$ in TensorRT. One crash is caused by an incompatible argument in dynamic tensor shape. One crash results from a failure in memory allocation. Scalpel achieves a 100\% valid model rate (VMR) and 100\% operator coverage (OC) on Autoware, along with 80\% component coverage (CC), demonstrating excellent model generation effectiveness and testing sufficiency. Additionally, Scalpel maintains high efficiency when testing Autoware, with an average of 16.63 seconds to generate each unique model and 864 seconds to trigger each unique bug. These experimental results demonstrate Scalpel's strong performance on Autoware, thus proving its excellent generalization across different autonomous driving systems.}

\DIFadd{
 \section{Threats to Validity}
}

\DIFaddend In this section, we discuss the threats to validity in our paper, including the internal threats, external threats, and construct threats.

\DIFaddbegin \DIFadd{The internal threats lie in the model component repository setup and the inconsistency oracle. To make our model component repository setup more representative, we have gathered all deployed DL models within Apollo and Autoware, and incorporated all model components utilized by these models into our model component repository. Furthermore, Scalpel's model component selection, mutation, and assembly are adjustable, enabling incorporating more model components into our repository. To guarantee the reliability of the inconsistency oracle, Scalpel's test oracle for detecting inconsistency bugs closely follows prior DL framework testing works (e.g., Muffin [12], Gandalf [24]). Unlike crashes, inconsistency bugs manifest solely in the final output tensors, making them hard to locate to specific model components, and thus difficult for developers to confirm and repair. Nevertheless, these inconsistency bugs remain widely focused on and detected by existing methods, as these bugs often indicate framework incompatibilities and can degrade performance when models are transferred among DL frameworks.
}\DIFaddend 

\DIFaddbegin \DIFadd{The external threats lie in the stochasticity. To mitigate stochastic effects, each experiment runs for six hours and executes thousands of rounds. No new bugs are found in the final hour, indicating excellent convergence.
}\DIFaddend 

\DIFaddbegin \DIFadd{The construct threats lie in applied automotive DL components. Our algorithm design is framework-agnostic, so it can be easily extended to DL components beyond autonomous driving domain by adding these model components into Scalpel's repository. Scalpel will automatically mutate these model components, assemble them into new test input models, and deploy generated test input models on DL frameworks under test to conduct differential testing.
}\DIFaddend 
 
\section{Related Works}
\label{related works}

Bugs in DL frameworks and compilers have garnered extensive attention in recent years. Many empirical studies on bugs in DL frameworks and compilers have been undertaken, laying the groundwork for the testing methods \cite{DLbug1,DLbug2,DLbug3,DLbug4,DLbug5,DLbug6,DLbug7,DLbug8}. Among them, some methods detect bugs in DL compilers (e.g., TVM \cite{tvm}). Representative works in this area include Neuri \cite{NeuRI}, NnSmith \cite{nnsmith}, TVMfuzz \cite{TVMfuzz}, and MTDLComp \cite{mtdlcomp}. These methods only take into account the syntactic and semantic characteristics of the DL compiler itself, without considering the potential constraints when the DL compiler interacts with external systems. Consequently, their generated test inputs cannot be successfully deployed into the autonomous driving systems. Another methods detect bugs in DL frameworks (e.g., PyTorch). Based on the differences in test inputs, these methods are primarily categorized into two types: interface-based testing and model-based testing. Interface-based testing only generates tensors to invoke each operator in DL framework. Among interface-based testing, Predoo \cite{predoo} designs tensor mutation rules to detect output precision error. FreeFuzz \cite{freefuzz} extracts insights from real-world code and model executions to conduct automated API-level fuzzing. DeepREL \cite{DeepREL} infers potential relationships between APIs based on their syntactic and semantic information, and generates test programs to invoke relational APIs. Duo \cite{duo} applies nine mutation rules to generate test input tensors, heuristically guided by genetic algorithms. EAGLE \cite{eagle} concludes and invokes equivalent operators in DL frameworks. Model-based testing generates both test input models and tensors. Among model-based testing, Cradle \cite{cradle} directly collects publicly available models to test DL frameworks. LEMON \cite{lemon} chooses publicly available model seeds, and designs multiple mutation rules for these seeds. Gandalf \cite{gandalf} applies context-free grammar and deep-Q network to generate new test input models. Audee \cite{audee} randomly generates DL models by combining DL frameworks' different parameter configurations.  Muffin \cite{muffin} firstly designs the connections between operators, and then specifies the types of each operator. DLLEN \cite{dllen} and TitanFuzz \cite{titanfuzz} applies large language model to generate DL models. However, the above DL framework testing methods are limited to processing single input and output tensors and are incapable of handling multi-modal data, leading to failure in testing automotive DL frameworks.

 \section{Conclusion}
\label{conclusion}

In this work, we propose Scalpel, an automotive DL framework testing method via assembling model components. Scalpel utilizes three model components (head, neck, and backbone) to process multiple input/output tensors, process multi-modal data, and extract data features at multiple levels. Additionally, Scalpel sets up a model component repository to provide available model components. From this repository, Scalpel selects, mutates, and assembles model components to generate DL models. We evaluate Scalpel in Apollo, and test Apollo's native automotive DL framework, PaddleInference. The experimental results show that Scalpel detects 16 crashes and 21 NaN \& inconsistency bugs. Furthermore, Scalpel attains improvements of 27.44$\times$ in model generation efficiency and 8.5$\times$ in bug detection efficiency. \DIFadd{Additionally, when extending to Autoware, Scalpel detects nine crashes and 16 NaN \& inconsistency bugs, which shows excellent generalization to other autonomous driving systems.} These results highlight the practical value of Scalpel in ensuring the quality of automotive DL frameworks, ultimately contributing to the safety and robustness of autonomous driving systems. In the future, we will focus on and exploring more optimizations for automotive deployment scenarios.

\section*{Acknowledgments}
We would like to thank the anonymous reviewers for their time and comments.
This work is partially supported by the National Key Research and Development Program of China (2024YFF0908004), the National Natural Science Foundation of China (U24A20337, 62372228), the Fundamental Research Funds for the Central Universities (14380029), and the Shenzhen-Hong Kong-Macau Technology Research Programme (Type C) (Grant No.SGDX20230821091559018).

\bibliographystyle{ACM-Reference-Format}

\bibliography{sample-base}

\end{document}